\documentclass[twocolumn,twoside]{article}
\usepackage{graphics}
\newcommand{\Teff}{${T_{\rm eff}}$ }

\newcommand{\hb}{\\ \hspace*{2ex}}

\begin{document}
\title{EFFECTIVE TEMPERATURE AND RADIAL VELOCITY OF THE SMALL--AMPLITUDE
CEPHEID POLARIS ($\alpha $ UMi) IN 2015}
\author{I.~A.\,Usenko,$^{1,2}$ V.~V. Kovtyukh,$^1$ A.~S.\,Miroshnichenko,$^3$ S.Danford$^3$\\[2mm]
\begin{tabular}{l}
 $^1$ Astronomical Observatory, Odessa National University, Shevchenko Park, \hb
  Odessa 65014, Ukraine, {\em igus99@ukr.net}\\
$^2$ Mykolaiv Astronomical Observatory, Mykolaiv 54030, Ukraine\\
$^3$ Dept. of Physics and Astronomy, University of North Carolina at Greensboro,\hb Greensboro, NC, USA  
\\[2mm]
\end{tabular}
}
\date{}
\maketitle

\indent

ABSTRACT.
We present the results of an analysis of 21 spectra
of $\alpha$ UMi (Polaris) obtained in September -- December
2015. Frequency analysis shows an increase of the pulsation period up
to 8.6 min in comparison to the 2007 observational set. The radial velocity
amplitude comes to 4.16 km\,s$^{-1}$, and it approximately twice the
one found in 2007. The average \Teff = 6017 K, and it is
close to the value determined for the 2001--2004 set. Therefore
Polaris moves to the red edge of the Cepheid instability strip (CIS).\\[1mm]

{\bf Key words}: Stars: radial velocities; Cepheids: overtone pulsations;
Cepheids: companions; Cepheids: $\alpha$ UMi\\[2mm]

{\bf 1. Introduction}\\[1mm]
Polaris is the nearest Cepheid to us in the Galaxy. Being a small-amplitude
pulsator (DCEPS) it has some specific features which testify about
its peculiar character:
\begin{enumerate}
\item An abrupt decrease of the pulsation amplitude during forty years
(5--6 km\,s$^{-1}$ before 1950 (Roemer 1965) to 0.05 km\,s$^{-1}$ in the
1980's (Fernie, Kamper, \& Seager 1993)) and the beginning of its
increase during the next 20 years (Dinshaw et al. 1987; Hatzes \& Cochran 2000;
Kamper 1996; Bruntt et al. 2008; Lee et al. 2008), -- from 1.5 km\,s$^{-1}$
in 1987 to 2.4 km\,s$^{-1}$ in 2007.

\item According to Turner et al. (2005), both photometry and radial velocity data taken
in 1896--2004 suggest an increase of the pulsation period by 4.45$\pm$0.03 s\,yr$^{-1}$
(from 3.966942 to 3.970691 days), except for a short-term slowdown of 4.28$\pm$0.73 s\,yr$^{-1}$
in 1963--1966. The period increase is an evidence of Polaris' redward crossing of the Cepheid
instability strip (hereafter CIS). According to Lee et al. (2008), the pulsation period increased
up to 86 seconds in 2005--2007 (from 3.973000 to 3.97394 days), while the pulsation
amplitude increased to 2.2 km\,s$^{-1}$.

\item Previous frequency analyses of the radial velocity data sets of 1987/88
(Dinshaw et al. 1989), 1992/93 (Hatzes \& Cochran 2000), Kamper (1996),
Lee et al. (2008) revealed the presence of additional periods of 45.3, 40.2, 34, and 119 days, respectively,
in addition to the main pulsation period of 3.97 days.
These additional periods have been explained by the rotation of Polaris,
existence of cool or macroturbulent velocity spots, or non-radial pulsations.\\[2mm]
\end{enumerate}

{\bf 2. Observations and frequency analysis}\\[1mm]
Twenty one spectrum were taken in September--December 2015 with the 0.81\,m
telescope of the Three College Observatory (TCO), located in central North
Carolina, USA. They were obtained with an \'echelle spectrograph manufactured
by Shelyak Instruments\footnote{http://www.shelyak.com} in a spectral range
from 4250 to 7800 \AA\ with a spectral resolving power of $R \sim$ 10000.
There are no gaps between the spectral orders. The data were reduced using
the {\it \'echelle} package in IRAF.

The DECH30 package (Galazutdinov 2007) allows to measure the line
depths and radial velocities using spectra in FITS format. Lines depths
were used to determine the effective temperature (a method based on
the spectroscopic criteria, Kovtyukh 2007).
Derived values of the \Teff\ and radial velocity for each spectrum are given
in Table 1.

\begin{table*}
\caption{Observational data of $\alpha$ UMi}
\small
\begin{tabular}{ccccrccc}
\noalign{\smallskip}\hline
    HJD   & \Teff & Phase    & \multicolumn{5}{c}{$RV$ (km\,s$^{-1}$)}\\
 2450000+ &    K       &       & Metals & NL    & H$_{\alpha}$ & H$_{\beta}$ & H$_{\gamma}$\\
\hline
7283.656  & 6083$\pm$15 & 0.854 & --17.57$\pm$1.48 & 149 & --19.27 & --17.18 & --17.20\\
7284.715  & 6059$\pm$23 & 0.120 & --19.10$\pm$1.63 & 139 & --20.12 & --17.79 & --17.73\\
7311.627  & 6048$\pm$20 & 0.890 & --11.93$\pm$1.38 &  104 & --11.27 & --10.04 & --10.93\\
7312.630  & 6064$\pm$20 & 0.142 & --18.73$\pm$1.71 & 114 & --19.50 & --17.67 & --16.97\\
7313.632  & 6007$\pm$15 & 0.394 & --15.00$\pm$1.59 & 129 & --16.72 & --15.51 & --15.20\\
7314.583  & 5997$\pm$16 & 0.633 & --15.83$\pm$1.19 & 119 & --16.43 & --14.96 & --15.93\\
7316.650  & 6070$\pm$17 & 0.153 & --18.19$\pm$1.72 & 121 & --19.31 & --17.13 & --17.44\\
7317.625  & 5967$\pm$16 & 0.398 & --15.52$\pm$1.56 &   91 & --16.44 & --14.48 & --14.81\\
7318.599  & 6001$\pm$18 & 0.643 & --14.93$\pm$1.29 & 118 & --15.57 & --14.18 & --15.27\\
7319.535  & 6077$\pm$16 & 0.879 & --17.36$\pm$1.36 & 118 & --18.90 & --16.85 & --17.11\\
7337.676  & 5962$\pm$17 & 0.442 & --16.08$\pm$1.08 & 132 & --17.69 & --15.21 & --16.11\\
7338.548  & 6006$\pm$19 & 0.661 & --15.35$\pm$1.67 & 125 & --16.09 & --14.61 & --14.35\\
7339.657  & 6025$\pm$20 & 0.940 & --18.54$\pm$1.45 & 141 & --20.36 & --17.74 & --17.90\\
7340.667  & 6037$\pm$15 & 0.194 & --17.65$\pm$1.35 & 134 & --19.13 & --16.87 & --17.36\\
7343.559  & 6058$\pm$18 & 0.922 & --18.55$\pm$3.36 & 139 & --20.42 & --18.33 & --17.77\\
7349.546  & 5990$\pm$14 & 0.428 & --15.15$\pm$1.37 & 138 & --16.73 & --14.34 & --15.01\\
7350.581  & 6034$\pm$16 & 0.688 & --14.43$\pm$1.62 & 123 & --15.59 & --15.06 & --14.38\\
7365.541  & 6025$\pm$17 & 0.451 & --15.14$\pm$1.47 & 147 & --15.71 & --14.34 & --14.72\\
7366.563  & 5973$\pm$19 & 0.708 & --15.20$\pm$1.52 & 121 & --15.94 & --14.26 & --14.42\\
7372.638  & 5975$\pm$15 & 0.236 & --17.76$\pm$1.48 & 154 & --19.27 & --17.18 & --17.20\\
7376.586  & 6022$\pm$14 & 0.229 & --16.60$\pm$1.34 & 161 & --18.27 & --15.92 & --12.20\\
\hline
\end{tabular}
\end{table*}

In the next step we used the PERIOD04 program (Lenz \& Breger 2005) to calculate a Fourier power spectrum. It
uses the Fourier and Fast Fourier Transform analysis with minimization of
residuals of sinusoidal fits to the data.

The power spectrum was calculated over the frequency range
0--1 d$^{-1}$ with a resolution of 0.00002 d$^{-1}$. The highest
amplitude of 1.70 corresponds to a frequency of 0.25126439 d$^{-1}$, or
3.97987156 days, respectively. This period exceeds by 8.5 minutes the period
of 3.97394 days determined from the 2007 observational set.
The systemic velocity ($\gamma$ -- velocity) is equal to $-$16.70
km\,s$^{-1}$.\\[2mm]

The following ephemeris has been computed based on the radial velocity values:

\begin{equation}
RV_{\rm min} = HJD\,2457284.237 + 3.97987156 \times E
\end{equation}

The radial velocity data for each spectrum folded with this period are given in Table 1.

The radial velocity and effective temperature of Polaris are shown in Figures 1 and 2
(data from the last three months of 2015).

The data were fitted with the sinusoidal curves.
However, one value of  $-$11.93$\pm$1.3 km\,s$^{-1}$ obtained on HJD 2457311.627 was
excluded from the analysis. According to this approximation, the mean amplitude
of the radial velocity curve is 4.16 km\,s$^{-1}$.\\[2mm]

\begin{figure}
\resizebox{\hsize}{!}{\includegraphics{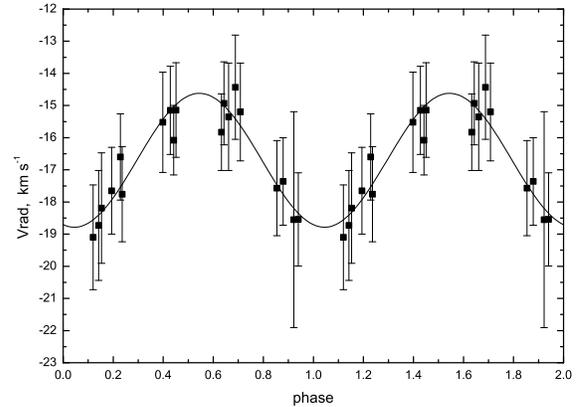}}
\caption{The radial velocity variations of Polaris during its pulsation period.}
\label{Fig1}
\end{figure}

\begin{figure}
\resizebox{\hsize}{!}{\includegraphics{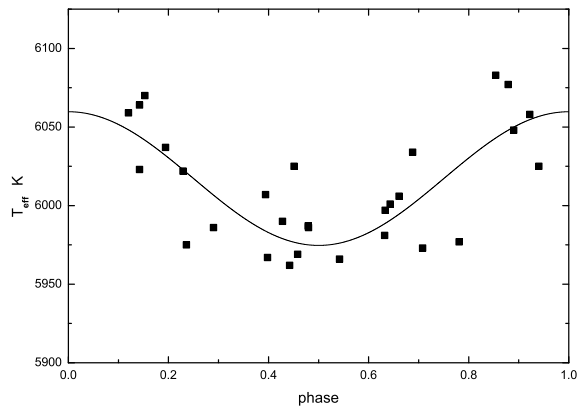}}
\caption{\Teff\ of Polaris during its pulsation period.}
\label{Fig2}
\end{figure}

{\bf 3. Summary}\\[1mm]
\begin{enumerate}
\item As seen from the results of our observations, the pulsation period of
Polaris shows a considerable increase in comparison with estimates obtained
in 2007 (Lee et al. 2008).  This fact confirms the assumption that Polaris
is moving to the red edge of the CIS;
\item The mean amplitude of the radial velocity nearly doubled during
the last eight years in comparison to the 2007 observations by Lee et al. (2008).
\item The effective temperature of Polaris for this data set averages
6017 K. This value is close to 6015 K determined for the
set of 2001--2004 data (Usenko et al. 2005).\\[2mm]
\end{enumerate}

{\bf References\\[2mm]}
Bruntt, H., Evans, N.R.,  Stello, D., Penny, A.J., Eaton, J.A., Buzasi, D.L.,
Sasselov, D.D., Preston, H.L. \& Miller-Ricci, E.: 2008, {\it ApJ} {\bf 683},
433\\
Dinshaw, N., Matthews, J.M., Walker G.A.J.\& Hill G.M.: 1989. {\it AJ,}
  {\bf 98}, 2249\\
Evans, N.R., Sasselov, D.D., \& Short, C.I.: 2002, {\it ApJ,} {\bf 567}, 1121\\
Fernie, J.D.: 1966, {\it AJ,} {\bf 71}, 731\\
Fernie, J.D., Kamper, K.W., \& Seager, S.: 1993, {\it ApJ,} {\bf 416}, 820\\
Galazutdinov, G.A.: 2007, {\it http://gazinur.com/DECH-software.html}\\
Hatzes, A.P., Cohran, W.D.: 2000, {\it AJ} {\bf 120}, 979\\
Kamper, K.W., Evans, N.R., and Lyons, R.W.: 1984, {\it JRASC,} {\bf 78}, 173\\
Kamper, K.W.: 1996, {\it JRASC,} {\bf 90}, 140\\
Kamper, K.W., Fernie, J.D.: 1998, {\it AJ,} {\bf 116}, 936\\
Kovtuykh, V.V.: 2007, {\it MNRAS,} {\bf 378}, 617\\
Lee, B.-C., Mkrtichian, D.E., Han, I., Park, M.-G., \& Kim, K.-M.:  2008,
{\it AJ,} {\bf 135}, 2240\\
Lenz, P. \& Breger, M.: 2005 {\it Commun. Astroseismology} {\bf 146}, 53\\
Roemer, E.: 1965, {\it ApJ,} {\bf 141}, 1415\\
Turner, D.G., Savoy, J., Derrah, J., Sabour, M.A., \& Berdnikov, L.N.:
2005,{\it AJ,} {\bf 117}, 207\\
Usenko, I.A., Miroshnichenko, A.S., Klochkova, V.G. \& Yushkin, M.V.:
2005, {\it MNRAS,} {\bf 362}, 1219\\

\end{document}